# What do you say? A pilot study investigating student responses in Data Driven Classroom Interviews

Jaclyn Ocumpaugh[1][0000-0002-9667-8523], Zhanlan Wei[2][0009-0002-3931-6398], Amanda Barany[2][0000-0003-2239-2271], Xiner Liu[2][0009-0004-3796-2251], Andres Felipe Zambrano[2][0000-0003-0692-1209], Ryan Baker[3][0000-0002-3051-3232], Camille Gioradno[2][0009-0001-2772-3079]

[1] University of Houston, Houston TX 77204, USA
[2] University of Pennsylvania, Philadelphia PA 19104, USA
[2] Adelaide University, Adelaide SA 5005, AUS.
jocumpau@cougarnet.uh.edu

**Abstract.** Data that contextualizes student interactions with online learning systems can be challenging to obtain. This study looks at the rhetorical strategies of a novel method for conducting in-the-moment Data-Driven Classroom Interviews (DDCIs). By using Ordered Network Analysis (ONA) to reanalyze data from Wei et al.'s (2025) Epistemic Network Analysis, we better account for the sequences in which these rhetorical strategies emerge during the interview process. Specifically, we examine how five rhetorical strategies by interviewers relate to five possible rhetorical strategies used in student responses. As with the previous study, results demonstrate minor differences in how students with high and low situational interest respond. Namely, whereas students with high situational interest show moderately higher levels of enthusiasm, students with low situational interest are more likely to respond to interviewers with an explanation. However, overall this study confirms that there are few interviewer-driven differences in these interviews, and it documents that interviewers are following guidelines to rely upon open-ended questions.

## 1 Introduction

Determining the best strategies for interviewing is complicated, especially when the interviewee is a child. Research in several fields suggests that it is best practice to let children do something they enjoy (such as drawing) while you interview them (Butler et al., 1995; Fargas et al., 2010), and research in several areas has worried about wording questions in ways that do not lead the child to one particular answer (Aidridge & Cameron, 1999; Davies et al., 2000; Lamb 2016; Powell & Guadagno, 2008). In informal education, research has looked at the effect of training parents to use *Wh*-questions with their kids when interacting with museum exhibits (Chandler-Campbell et al., 2020). In classroom research, studies examining how kids respond to questions have found cultural differences in who responds with narratives versus who responds with facts (Heath, 1982; 2012).

Rhetorical advice suggests interviewers should use open-ended questions over closed-ended questions (e.g., Cederborg & Lamb, 2008; Lamb et al., 2011) and wait longer for responses than they would in a normal conversation (Shiau et al., 2024), but overall little work has examined interviewers' rhetorical strategies, particularly in a quantitative fashion. The quantitative studies that do exist are primarily for forensic purposes (e.g., Brown et al., 2013; Dent & Stephenson, 1979; Lamb et al., 1996). They have found that training can improve the quality of information that law enforcement obtains from child witnesses (Schreiber Compo et al., 2012). However, they have also found that interviewers often do not follow best practice guidelines, including those that recommend the use of open-ended questions (Cederborg & Lamb, 2008).

Forensic interviews can be very different than a research-driven interview where (a) it may not be as necessary to ensure that children perfectly recall high-stakes events and (b) the important steps that forensic interviewers take to avoid retraumatizing a child are (we hope) not necessary in most research-based interviews in education. In fact, education researchers might expect that children's memory of a learning activity might differ from actual, recorded events, and researchers may benefit from documenting those discrepancies. As such, researchers might want to use different rhetorical strategies than those employed in forensic tasks, and we would benefit from understanding which of our own rhetorical moves are common and/or effective for a given research context.

In this study we seek to address this gap by investigating the rhetorical strategies in a novel interview method call Data Driven Classroom Interviews (DDCIs; Baker et al., 2023; Ocumpaugh et al., 2025), which are facilitated by an app called Quick Red Fox (QRF; Hutt et al., 2022), which signals the researcher whenever an event of interest occurs within an online learning software. Specifically, we use ordered network analysis (ONA; Tan et al., 2023) to investigate how the rhetorical strategies of interviewers manifest during DDCIs and the degree to which these strategies potentially influence the rhetorical strategies of students. These interviews which were conducted in the context of understanding the development of students' STEM interest as they engaged with WHIMC (What-If Hypothetical Implementations in Minecraft)—an immersive, game-based platform designed to cultivate students' interest in STEM (Lane et al., 2017). Therefore, in addition to examining which patterns of question and response types were most common, we also consider differences in students' situational interest (Linnenbrink-Garcia et al., 2010).

## 2  Related Work

### 2.1  Interview approaches

The use of interviews in education research is long established, and has been particularly important for areas of research like interest development, where it may be harder for observational protocols to capture the construct under investigation. For example, Lipstein

& Renninger (2007) used interviews to better understand differences between the early phases of interest development. It has also been used to investigate issues like changes in conceptual reasoning (Morales-Navarro et al., 2025).

Typically, these interviews have been conducted before or after learning experiences, although Think Aloud (TA) protocols—where students are asked to talk during their use of an online learning system, for example—have sometimes also been used as well (Ericsson & Simon, 1993). TAs have an advantage of allowing us to capture things in the moment, which can be important for understanding concepts like struggle, which may be remembered differently depending on whether or not the student was able to resolve their impasse. However, they may also increase the cognitive load on the student during the learning experience, and students may need to be prompted to continue talking or to stay on topic (see discussion in Ocumpaugh et al., 2025).

Recently, a more adaptive interview technique—Data Driven Classroom Interviews (DDCIs)—has been introduced to help researchers access student thinking as it evolves in real time (Ocumpaugh et al., 2025). DDCIs are designed to facilitate brief, strategically timed interviews during active engagement in learning environments where students are immersed in interactive tasks. DDCIs are supported by a tool called Quick Red Fox (QRF; Hutt et al., 2022), an open-source application that detects pre-specified indicators of interest-related behaviors and informs interviewers to engage with students at relevant moments. By guiding timely conversations at key points during learning, QRF enables researchers to gather rich, contextualized data on students' perceptions, affective engagement, and reflections as they occur. In doing so, the system supports researchers in delivering timely, targeted responses aligned with students' emerging learning needs.

### 2.2 Analyzing Interviews With Epistemic Network Analysis

Epistemic Network Analysis (ENA; Shaffer et al., 2017) has shown promise in helping researchers to better understand interview data. For example, Cheung and Winterbottom (2023) analyzed semi-structured interviews, using ENA to examine how students reported learning from visual representations in textbooks. Likewise, Orrill and Brown (2024) analyzed think-aloud and interview data to better understand teacher strategies. These studies align with ENA's goals to model the structure of connections among coded elements in discourse data (Shaffer, 2017), but to date, this work is still relatively nascent and has focused more on content than on speech acts or other rhetorical components.

### 2.3 Situational Interest

Theories of interest development suggest that internalized, individual interest requires the acquisition of domain-specific knowledge, and until then students engagement is driven primarily through situational interest (Hidi & Renninger, 2006). This latter form of interest is characterized as short-term and context-dependent, but nonetheless influential (Linnenbrink-Garcia, 2010). Because of its importance in knowledge acquisition (Hidi & Renninger, 2006), research has examined how situational interest is triggered during learning activities (Palmer, 2009) and how it supports students in directing attention (Linnenbrink-Garcia et al., 2013), sustaining cognitive effort (Thoman et al., 2011), and building the foundational understanding needed to develop more lasting individual interest (Hidi & Renninger, 2006). In some studies, researchers have also looked at how situational interest affects students' in-game language usage (Wei et al., 2025), but to date, little research has examined how students describe their experiences more directly. In this study we begin that effort by examining broadly how the relationship between interviewer questions and student response types. Specifically we look at the dynamics of question types and response types to better understand the effects of an interviewer's rhetorical strategies on the ways in which a student might respond during a DDCI.

## 3 Methods

### 3.1 Learning System, Participants, and Survey Measures

Data was collected from a five-day, summer camp in the northeastern US in 2024. Students interacted with WHIMC, a simulation constructed using Minecraft's Java Edition where students are encouraged to explore immersive astronomy-related scenarios (Lane et al., 2017; Lane et al., 2022; Yi et al., 2021). In total, 14 students participated: 10 male, 3 female, and 1 non-binary; 9 White/White Americans, 2 Asian Americans, and 3 students who preferred not to answer. At the start of camp, assent and informed consent were obtained from both students and their guardians. On day four, students took Linnenbrink-Garcia et al.'s (2010) situational interest (SI) measure, which was used to categorize students into high, low, and medium SI using the median and standard deviations. In this pilot study, we compare the High-SI (n=6) and Low-SI (n=4) groups in order to examine how students with differing levels of interest respond during DDCIs. Notably, this is a re-analysis of Wei et al.'s (2025) data that used the same coding scheme. However, the comparison in this study uses a different statistical analysis (described in section 3.4) to help us better explore the degree to which student response types may have been induced by certain interview strategies.

### 3.2 Data Driven Classroom Interviews

During the five-day camp, two researchers conducted data driven classroom interviews (DDCIs; Baker et al., 2023; Ocumpaugh et al., 2025). DDCIs constitute a new interview method that is facilitated by an open-source app called the Quick Red Fox (QRF; Hutt et al., 2022; Ocumpaugh et al., 2025). In this method, the QRF server places predetermined events related to a research question into a prioritization queue so that each time a researcher signals to the app that they are ready for a new interview, the QRF app sends an interview trigger that reflects a recent, high-priority event. The researcher can then approach that student in the moment, while the event is still relevant to the student's experience. The QRF app allows the researcher to record the interview directly within the app, which aggregates all associated metadata for future analysis.

**DDCI Preparation.** In this pilot study, we use DDCIs to investigate student interest development, so the events used to trigger interviews were related to behaviors that indicated high and low engagement within WHIMC. However, starting an interview by asking students about the triggering event (i.e., "Why did you just do X?") is likely to make them feel overly surveilled and defensive. Moreover, the typical DDCI is relatively short compared to other interview methods (in many cases 3-5 minutes each). Although the researcher will interview the same students many times over the course of the fieldwork, it is important to work in ways that build rapport over time. As such, pre-fieldwork efforts included the development of a range of questions that might help to (a) build rapport with students (b) encourage students to report accurately about their gameplay experience, and (c) facilitate open conversations about students interests both within the game and beyond. (See discussion in Ocumpaugh et al., 2025).

**DDCI Implementation.** In this case, students interact with WHIMC over the course of one week summer camp, during which two interviewers conducted DDCIs. As with previous research, any interviews were in the 3-5 minute range, but some went as long as 19 minutes. Over the course of the week each of the 14 students was interviewed 9 to 19 times.

In keeping with the goals of the DDCI method, WHIMC students typically continued to play the game during each interview. This method allows interviewers to observe how the students' are responding emotionally to the learning experience as they shape their questions for the interview. In this study, interviewers focused on trying to understand how students' interactions with the WHIIMC system related to their science interest development, as that was the underling goal of the research project.

However, this data also allows us a opportunity to study the interviewer—which is often assumed to be a constant rather than a variable in interview-based research. In particular, we are able to examine the rhetorical strategies of the interviews in this project, in order to determine if they are following the advice about how to approach students as outlined in Ocumpaugh et al., (2025). Moreover, we are able to determine whether these strategies are responsible for the minor differences exhibited by students with high and low situational interest, as documented by Wei et al., (2025).

### 3.3 Transcription and Coding

Interviews were deidentified during the transcription process, which closely mirrored an established protocol for representing conversational data, including features like pauses, false starts, and incomplete sentences (PLDC, 2011). Data was then coded to better understand the degree to which interviewers' rhetorical strategies influence student responses. For this preliminary investigation two sets of codes were developed to categorize (1) the interviewers' questions (i.e., *Open-ended Close-ended, Follow-up, Process,* and *Reframing*; see Table 1) and (2) the students' responses (e.g., *Explanation, Brief, Enthusiastic, Neutral,* and *Redirect*; see Table 2). These codes are not mutually exclusive; for example, a student response, could be both *Brief* and *Enthusiastic*.

Table 1. Final Codebook for Interviewer (I) Questions to Students (S)

| Question Types | Examples |
| --- | --- |
| *Open-Ended:* asks students to provide an elaborative or reflective response rather than a brief or one-word answer. | **I:** *What do you like about it?* |
| *Close-ended:* asks students to provide a brief, specific response, often limited to a single word of predefined options, such as 'YES' or 'NO'; typically seeks factual answers or confirmations rather than encouraging further explanation. | **I:** *Were you able to place objects before?* |
| *Follow-up:* asks students to expand on their initial responses, clarify their reasoning, or delve into their thought processes more deeply. | **I:** *Did he morph into an owl?* <br> **S:** *Yes, he is Duolingo.* <br> **I:** *What is he doing as Duolingo?* |
| *Process:* asks students to describe their thought process, actions, decision-making, or strategy use while engaging in a task. | **I:** *So, what's your strategy for blowing up Saturn?* |
| *Reframing:* restates part of the student's original statement, either by repeating a key phrase, rewording the sentence, or echoing the last word with a clarifying tone to check understanding. | **I:** *Are the days warmer?* <br> **S:** *There are no days, it's ...* <br> **I:** *Oh, there's no days. Ohhh.* |

Table 2. Final Codebook for Student (S) Responses to Interviewer (I) Questions

| Response Types | Examples |
| --- | --- |
| *Explanation:* responses that provide reasoning, clarification, or a detailed description of their thinking, actions, or decision-making process. | **I:** *Why did you decide to put that deeper underground?* <br> **S:** *Because I wanted to, and there's less radiation. So, you don't want radiation exposed. I mean that's why I'm wearing a hazmat suit.* |
| *Brief:* short replies, usually containing min. elaboration or details; often consists of a single word/short phrase that directly addresses the question without expanding on the thought process or providing additional context. | **I:** *That armor; have you seen that before?* <br> **S:** *Nah* |
| *Enthusiastic:* conveys excitement, joy, or is associated with other positive reactions. Often includes playful language, expressive tones, or positive affective markers. | **I:** *Snow?* <br> **S:** *And a pressure plate. Whee!* |
| *Neutral:* does not clearly reflect enthusiasm/is short in length; often serves to confirm detail or make a factual statement w/o signaling interest or disinterest; typically moderate in tone/lacking strong emotional expressions. | **I:** *What material is that?* <br> **S:** *It's black concrete.* |
| *Redirect:* shifts focus from the original question asked by the interviewer instead of directly addressing the inquiry; often unrelated/only loosely connected to questions asked. | **I:** *Like, what separates a computer from a robot?* <br> **S:** *Okay, I'm starting the underground section.* |

**Codebook application and IRR.** Codes are not mutually exclusive, but were applied in a binary fashion at the utterance level, as documented in the transcriptions. After an initial round of coding, interrater reliability (IRR) was calculated between two members of the research team, who then addressed disparities through social moderation (Herrenkohl & Cornelius, 2013) before reassessing IRR with a new subset of the data. Once IRR was established by the human raters, codes was deemed acceptable for automation and GPT-4o was utilized to automate the coding process, following a few-shot prompting approach, where we fed GPT-4o with the name, definition, and selected examples of each code. (Xiao et al., 2023; or Liu et al. 2025). However, in the case of four codes (*Follow-up, Process, Reframing*, and *Redirect*), GPT-4o was unable to reach an adequate IRR with human coders (i.e., κ>0.7) and coding was completed by a human coder who had reached IRR (Table 3). To further ensure GPT-4o's reliability, each data set was processed three times under identical configurations, and a majority vote labeled the data.

Table 3. Inter-rater reliability between human-human and human-GPT.

| Category | Codes | Human-Human | | Final Coding Approach | Human-GPT | | |
|---|---|---|---|---|---|---|---|
| | | Initial κ | Final κ | | κ | Precision | Recall |
| Interview Questions | Open-ended | 0.89 | - | GPT-4o | **0.82** | 1.00 | 0.73 |
| | Close-ended | 0.81 | - | GPT-4o | **0.75** | 0.83 | 0.71 |
| | Follow-up | **0.78** | - | Manual | 0.57 | 0.45 | 1.00 |
| | Process | 0.25 | **0.78** | Manual | 0.50 | 0.44 | 0.78 |
| | Reframing | **0.75** | - | Manual | 0.47 | 0.38 | 0.75 |
| Student Responses | Explanation | 0.82 | - | GPT-4o | **0.78** | 0.73 | 0.89 |
| | Brief | 0.80 | - | GPT-4o | **0.79** | 0.79 | 0.93 |
| | Enthusiastic | 0.58 | 0.89 | GPT-4o | **0.75** | 0.71 | 0.83 |
| | Neutral | 0.67 | 0.72 | GPT-4o | **0.78** | 0.86 | 0.86 |
| | Redirect | **0.76** | - | Manual | 0.39 | 0.35 | 0.82 |

### 3.4 Ordered Network Analysis

As noted in Section 3.1, this paper is a reanalysis of Wei et al.'s (2025) work, which employed Epistemic Network Analysis to explore this data. In this reanalysis, we use a similar but distinct method: Ordered Network Analysis (ONA; Tan et al., 2023), which captures the directionality of the links between different discourse events. That is, where as ENA would treat a process question followed by a brief response as identical to a brief response followed by a process question, ONA captures the sequential order of the data by preserving the directionality of each connection.

To examine relationships among the various rhetorical strategies, we implemented ONA with the ENA Web Tool (V.1.7.0; Marquart et al., 2021). Student interest level (high SI vs. low SI) serves as the primary unit variable, while their unique player ID—which is nested under each interview—serves both as the secondary unit variable and the conversation variable. This structure reflects the nature of DDCI data, where students are interviewed multiple times over the course of a week, with some disconnection between individual interviews with each student. We apply an infinite stanza window, which treats all interactions in an interview as a continuous discourse event. Both interviewer questions and student responses are included in the ONA model in order to capture the cross-role dialogue between researchers and students.

## 4 Results

### 4.1 Overview and Descriptives

This study analyzes data from 10 of the 14 students in the pilot study. (Table 4) presents a breakdown of how the different types of questions and response were distributed, per student, across the 70 interviews being analyzed. The reader should note that because the codes were not mutually exclusive (i.e., a question could receive more than one label, as could a response), there are approximately 3 times as many response labels as questions. As this breakdown shows, interviewers approached the two groups of students in a very similar fashion. Although students with high situational interest received slightly higher numbers of questions overall, students with high and low situational interest received roughly the same distribution of question types. This finding is supported by a Mann-Whitney test showed no statistically significant differences between the high- and low-SI groups along the x-axis (U=3.00, p=0.11, r=0.70 at alpha=0.05).

**Table 4.** Distribution of Question and Response Types Averaged Per Student Across Interviews

| Type | Codes | High SI (N-6) | | Low SI (N=4) | | Dif | |
|---|---|---|---|---|---|---|---|
| | | N | % | N | % | N | % |
| Questions | Open-ended | 72.6 | 9% | 71 | 11% | 1.6 | -2% |
| | Close-ended | 31.2 | 4% | 19.3 | 3% | 11.9 | 1% |
| | Process | 63.8 | 8% | 56.3 | 9% | 7.5 | -1% |
| | Follow-up | 46 | 6% | 37 | 6% | 9 | 0% |
| | Reframing | 26.6 | 3% | 19.8 | 3% | 6.8 | 0% |
| Responses | Enthusiastic | 56 | 6% | 21.3 | 3% | 34.7 | 3% |
| | Brief | 194.4 | 24% | 144.8 | 21% | 49.6 | 3% |
| | Neutral | 239.8 | 29% | 209.3 | 31% | 30.5 | -2% |
| | Explanation | 74.4 | 9% | 72 | 11% | 2.4 | -2% |
| | Redirect | 21.2 | 2% | 16.8 | 2% | 4.4 | 0% |

We next examine the ONA results. Figure 1 presents the these for High-SI and Low-SI students separately, while Figure 2 presents the difference model comparing these two groups. As the visualizations in Figure 1 show, *Neutral* responses are the largest node for both groups of students, followed by *Brief* responses and then *Open-ended* and *Process* questions.

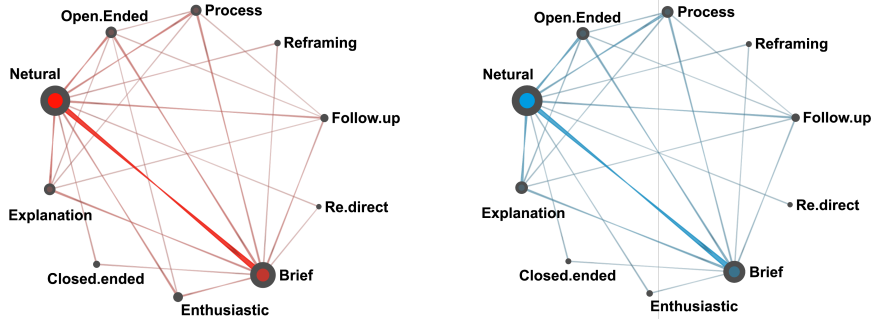

**Figure 1.** ONA Models for students with high and low situational interest. High-SI students are shown in red (left) and Low-SI students are shown in blue (right). Only LWs>0.03 are shown.

Although the similarities between the two groups of students are important, the visualization in Figure 2 shows a small but important number of transitions that are distinct enough to emerge in the difference model. These can primarily be seen in the students' responses, as only two nodes involving questions appear in this difference model. Students with low situational interest (blue) are more likely to respond to open ended questions with either a neutral response or an explanation, and they are more likely to respond to a process question with a neutral response. As Table 5 shows, the most common initial response from any student was *Neutral* or *Brief*, regardless of question type. This pattern can be found within both groups of students, regardless of their level of interest. Note, for example, that the transitions from *Open-Ended→Neutral* (LW=0.15 for High-SI; LW=0.19 for Low-SI) have line weights that are more than twice those for *Open-Ended→Explanation* or from *Open-Ended→Enthusiastic* (LWs=0.04 to 0.07).

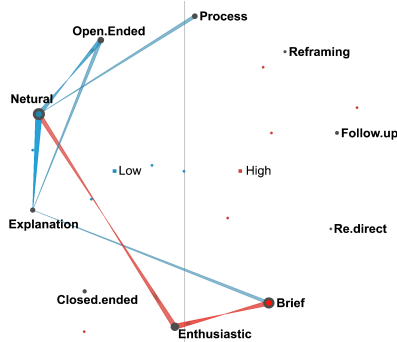

**Figure 2.** LW>0.03 for an ONA Difference Model comparing High-SI and Low-SI students.

### 4.2 Transitions from Interviewer to Student and then within the Student's Turn

However, as Table 5 shows, there are a few differences between the High and Low SI groups that are noteworthy. For example, High-SI students are nearly twice as likely to give *Brief* or *Neutral* responses to *Close-ended* Questions than Low-SI students, whose response types are dispersed among categories that did not make the threshold for reporting here (i.e., LW<0.03). High-SI students are also more likely to give *Enthusiastic* responses to *Open-Ended* questions than Low-SI students (LW=0.04 vs. LW=0.02). Interestingly, *Open-ended→Explanation* is the one transition occurring more often for Low-SI students than for High-SI students (LW=0.07 vs. LW=0.04).

**Table 5.** Transitions from Interviewer to Initial Student Response

| Question | Response | High-SI | Low-SI | Difference |
|---|---|---|---|---|
| Follow Up | Explanation | 0.03 | 0.03 | 0 |
| **Open-ended** | **Explanation** | **0.04** | **0.07** | **-0.03** |
| **Open-ended** | **Enthusiastic** | **0.04** | **0.02** | **0.02** |
| Reframing | Neutral | 0.06 | 0.05 | 0.01 |
| **Close-ended** | **Brief** | **0.06** | **0.03** | **0.03** |
| Follow Up | Brief | 0.07 | 0.07 | 0 |
| **Close-ended** | **Neutral** | **0.07** | **0.04** | **0.03** |
| Process | Brief | 0.10 | 0.09 | 0.01 |
| Follow Up | Neutral | 0.11 | 0.10 | 0.01 |
| Open-ended | Brief | 0.13 | 0.13 | 0 |
| Process | Neutral | 0.13 | 0.13 | 0 |
| Open-ended | Neutral | 0.15 | 0.19 | -0.04 |

*Note:* Data is sorted on the LWs for students with High-SI, bold face is used to indicate transitions where the LW of one group is at least twice that of the other.

After their initial response, students often provide other utterances before the turn shifts back to the interviewer. Results for these transitions are shown in Table 6, which also shows that transitions involving *Neutral* responses are most common. In fact, the most common transition in this data set is *Neutral→Neutral*, regardless of interest level (LW=0.47 for High-SI; LW=0.53 for Low-SI). Conversely, transitions involving *Brief*, *Redirect*, or *Enthusiastic* responses are less common. In fact, the only transitions involving a *Neutral* response with a LW<0.10 for High-SI students (or LW<0.06 for Low-SI students), are those that involve transitions to or from *Redirect* responses.

**Table 6.** Transitions from Response to Response

| Response | Response | High-SI | Low-SI | Difference |
|---|---|---|---|---|
| Redirect | Redirect | 0 | 0 | 0 |
| Neutral | Redirect | 0.03 | 0.04 | -0.01 |
| **Enthusiastic** | **Enthusiastic** | **0.03** | **0.01** | **0.02** |
| **Redirect** | **Brief** | **0.04** | **0.02** | **0.02** |
| Redirect | Neutral | 0.04 | 0.03 | 0.01 |
| **Reframing** | **Brief** | **0.05** | **0.03** | **0.02** |
| Explanation | Explanation | 0.05 | 0.07 | -0.02 |
| **Brief** | **Enthusiastic** | **0.09** | **0.04** | **0.05** |
| **Enthusiastic** | **Brief** | **0.09** | **0.04** | **0.05** |
| **Neutral** | **Enthusiastic** | **0.10** | **0.06** | **0.04** |
| Explanation | Brief | 0.11 | 0.14 | -0.03 |
| Brief | Explanation | 0.11 | 0.11 | 0 |
| **Enthusiastic** | **Neutral** | **0.11** | **0.05** | **0.06** |
| Explanation | Neutral | 0.14 | 0.21 | -0.07 |
| Neutral | Explanation | 0.15 | 0.18 | -0.03 |
| Brief | Brief | 0.32 | 0.24 | 0.08 |
| Neutral | Brief | 0.38 | 0.37 | 0.01 |
| Brief | Neutral | 0.39 | 0.35 | 0.04 |
| Neutral | Neutral | 0.47 | 0.53 | -0.06 |

*Note:* Data is sorted on the LWs for students with High-SI, bold face is used to indicate transitions where the LW of one group is at least twice that of the other

Transitions involving *Enthusiastic* and *Redirect* are also responsible for major differences between high and low SI groups. Here, the trend shows that these differences reflect a pattern where transitions from response to response are more likely among the High-SI students than the Low-SI students. For example, the line weights for *Enthusiastic→Neutral* were more than twice as high for the High-SI students than for the Low-SI students (LW=0.11 vs. LW=0.05), and the transition from *Redirect→Brief* only emerged among the High-SI students (LW=0.04 vs. LW=0.02).

### 4.3 Transitions from Student to Interviewers

We next look at the transitions back to the interviewer (e.g., after the student is done responding). Table 7 shows that the pattern for *Neutral* responses driving the strongest transitions holds true in this context as well, which is perhaps not surprising since it is the most common response type. Likewise, *Process* and *Open-ended* questions were also responsible for the most common transitions, with most line weights involving those question types showing up at LW>0.11.

**Table 7.** Transitions from Student back to Interviewer

| Response | Question | High-SI | Low-SI | Difference |
|---|---|---|---|---|
| Explanation | Follow up | 0.03 | 0.04 | -0.01 |
| Follow Up | Process | 0.03 | 0.03 | 0 |
| **Explanation** | **Open-ended** | **0.04** | **0.07** | **-0.03** |
| Brief | Reframing | 0.04 | 0.03 | 0.01 |
| Neutral | Reframing | 0.05 | 0.05 | 0 |
| **Brief** | **Close-ended** | **0.05** | **0.03** | **0.02** |
| Brief | Follow up | 0.06 | 0.06 | 0 |
| Neutral | Close-ended | 0.07 | 0.05 | 0.02 |
| Neutral | Follow up | 0.08 | 0.10 | -0.02 |
| Brief | Process | 0.11 | 0.11 | 0 |
| Neutral | Process | 0.12 | 0.16 | -0.04 |
| Brief | Open-ended | 0.12 | 0.12 | 0 |
| Neutral | Open-ended | 0.14 | 0.19 | -0.05 |
| Neutral | Neutral | 0.47 | 0.53 | -0.06 |

*Note:* Data is sorted on the LWs for students with High-SI, bold face is used to indicate transitions where the LW of one group is at least twice that of the other

There were few differences between the High-SI and Low-SI students in these transitions. *Explanation→Open-Ended* was more likely during interviews with Low-SI students (LW=0.07) than with High-SI students (LW=0.04). Conversely, only the transition from *Brief→Close-ended* was substantially more common in interviews with the High-SI group than the Low-SI group.

### 4.4 Transitions from Interview Question to Interview Question

Finally, we examined transitions from one question to another, which may sometimes be indicative of times that an interviewer did not get up-take on their original question, but may also indicate other times that the interviewer was attempting to switch strategies to better facilitate the interview. Not surprisingly, the strongest transitions in Table 8 are those that involve *Open-ended* or *Process* questions. This was consistent across High and Low-SI students, where *Open-ended→Open-ended* was the most common transition (LW=0.05 and LW=0.07). Only relatively minor differences were seen between the two groups, but *Open-Ended→Follow-up* only occurred for Low-SI students (LW=0.03), while *Reframing→Reframing* and *Close-ended→Close-ended*, while infrequent among High-SI students (LWs=0.01 for both transitions), did not occur in interviews with Low-SI students.

Table 8. Transitions from Question to Question

| Question | Question | High-SI | Low-SI | Difference |
|---|---|---|---|---|
| **Open-ended** | **Follow up** | **0** | **0.03** | **-0.03** |
| **Reframing** | **Reframing** | **0.01** | **0** | **0.01** |
| **Close-ended** | **Close-ended** | **0.01** | **0** | **0.01** |
| Follow Up | Follow up | 0.02 | 0.02 | 0 |
| Open-ended | Process | 0.04 | 0.06 | -0.02 |
| Process | Open-ended | 0.04 | 0.05 | -0.01 |
| Follow Up | Open-ended | 0.04 | 0.04 | 0 |
| Process | Process | 0.04 | 0.03 | 0.01 |
| Open-ended | Open-ended | 0.05 | 0.07 | -0.02 |

*Note:* Data is sorted on the LWs for students with High-SI, bold face is used to indicate transitions where the LW of one group is at least twice that of the other or where only one LW=0.00.

Notably, all LW differences were relatively small, but one place where High and Low-SI students diverge is in the transition from *Open-ended→Follow up* questions. Interviewers only made this transition when conducting DDCIs with Low-SI speakers (LW=0.03). This follows a general pattern among transitions involving two questionsk where the Low-SI students are slightly more likely to receive *Open-Ended* questions. Likewise *Reframing→Reframing* and *Closed-Ende→Close-Ended* were only used with High-SI students (LWs=0.01).

## 5 Discussion and Conclusions

### 5.1 Overview of this study

This study sought to document how rhetorical strategies used by an interviewer influence students' responses, as question types are often discussed in methodological papers but rarely quantified in studies of interviews. In addition, we sought to examine the degree to which students' situational interest in the task they were completing influenced their responses. Specifically, we sought to explore whether differences that emerged in an earlier analysis of this data, where students with low situational interest were more likely to give explanations (Wei et al., 2025), held up when data was using an Ordered Network Analysis, which distinguishes between *Open-Ended→Explanation* transitions and *Explanation→Open-Ended* transitions.

### 5.2 Important Similarities in the Data of High and Low-SI Students

Overall, the results show that both High and Low-SI students are likely to keep talking for at least two utterances, generating the strongest line weights for transitions from one student response to the next (Table 6). These utterances are most likely to include at least one *Neutral* response and may also include *Brief* responses, regardless of the student's interest level.

*Neutral* answers are the most likely response to most question types. When a student's turn ends, interviewers were most likely to transition to *Open-Ended* or *Process* questions, regardless of students' interest levels. This result is perhaps not surprising, as interviewers had been trained to privilege these rhetorical strategies, but it was important to see that both High-SI and Low-SI students were given the same opportunities to express themselves, particularly given research that suggests that interviewers sometimes abandon best practice (Cederborg & Lamb, 2008).

In cases where interviewers asked two questions in a row, transitions often involved *Open-ended* or *Process* questions. This occurred regardless of students' interest level, although LWs for *Open-ended* transitions were typically slightly higher for Low-SI students.

## 5.3   Important Differences in the Data of High and Low-SI Students

Most differences between High and LowSI students were small (Diff = -0.04 to 0.08). Notably, when looking at the responses to *Open-ended* questions, we see that High-SI students are equally as likely to give an *Enthusiastic* response as an *Explanation* (LW=0.04 for both transitions). However, Low-SI students diverged in terms of these two transitions, and were substantially more likely to give an *Explanation* (LW=0.07) than an *Enthusiastic* response (LW=0.02). This may reflect both a general pattern of engagement (e.g., lower enthusiasm among Low-SI students) as well as a higher need for metacognitive support. That is, when we ask Low-SI students about a question, they may need to verbally process what they are doing. This interpretation is supported by the *Open-Ended→Follow-up* findings that show Low-SI students (LW=0.03) are more likely to receive a *Follow-up* question after an *Open-ended* question than High-SI Students (LW=0.00). They are also slightly more likely to receive an *Open-ended* or a *Process* question after an *Open-ended* question, though the differences here are relatively weaker.

## 5.4   Limitations of this Study

One limitation of this study includes the combination of both rhetorical and engagement categories for student responses. Although both ENA and ONA allow for the analysis of codes that are not mutually exclusive, future research should consider models that consider these issues separately. This could include simply excluding either the rhetorical or engagement-related responses codes, but it could also be worth considering alternative ways to reconcile these analyses. For example recent work in transmodal analysis (Shaffer, Wang, & Ruis, 2025) might help us to differentiate between these typologies without excluding one of these categories. Likewise, future research might also consider other typologies, including, for example content-related categories that could be applied to both the question typologies and the response typologies.

## 5.5 Significance of this study

The current study investigates rhetorical strategies by both interviewers and their interviewees in a novel method for obtaining qualitative research fast, namely Data Driven Classroom Interviews (DDCIs). In it, we examine the effect of different question types on student response patterns in order to ensure that there are relatively few differences in how students of varying interest levels respond. The particular research question necessitates the use of a relatively generic typology of rhetorical strategies and response types, which may have lent itself to over-representing commonalities between High and Low-SI groups, who would likely diverge if the codes used for this ONA resulted from a thematic analysis based on the topics most closely related to situational interest.

That said, this pilot study demonstrates that situational interest is not significantly impacting either the rhetorical strategies that interviewers are applying to these students or the types of rhetorical moves they respond with. This kind of confirmation is important when ensuring that students' opportunities to talk are comparatively equal relative to the variable under investigation, but it leaves open further questions for future work, including those related to themes that may emerge more inductively in the data. It also raises questions about how these patterns might differ across different groups of students, and whether changing the interview strategies for students with lower motivational measures might provide support for the kinds of metacognition that could lead to improvements in learning and motivation.

**Acknowledgments.** This study was funded by NSF ECR #2301173. The first author is also grateful for the support from the 2024-2025 QE Fellows Institute at the Center for Research on Complex Thinking at the University of Wisconsin.

**Disclosure of Interests.** We have no conflicts of interest to disclose.